# Experimental Investigation on the Friction-induced Vibration with Periodic Characteristics in a Running-in Process under Lubrication

Di Sun [b], Pengfei Xing [a], Guobin Li [a,*], Hongtao Gao [c], Sifan Yang [a],

Honglin Gao [a], Hongpeng Zhang [a]

[a] Marine Engineering College, Dalian Maritime University, Dalian 116026, P. R. China

[b] Marine Engineering College, Jimei University, Xiamen 361021, P. R. China

[c] Naval Architecture and Ocean Engineering College, Dalian Maritime University, Dalian 116026, P. R. China

*Corresponding author: guobinli88@163.com

**Abstract:** This paper investigated the friction-induced vibration (*FIV*) behavior under the running-in process with oil lubrication. The *FIV* signal with periodic characteristics under lubrication was identified with the help of the squeal signal induced in an oil-free wear experiment and then extracted by the harmonic wavelet packet transform (*HWPT*). The variation of the *FIV* signal from running-in wear stage to steady wear stage was studied by its root mean square (*RMS*) values. The result indicates that the time-frequency characteristics of the *FIV* signals evolve with the wear process and can reflect the wear stages of the friction pairs. The *RMS* evolvement of the *FIV* signal is in the same trend to the composite surface roughness and demonstrates that the friction pair goes through the running-in wear stage and the steady wear stage. Therefore, the *FIV* signal with periodic characteristics can describe the evolvement of the running-in process and distinguish the running-in wear stage and the stable wear stage of the friction pair.

# 1. Introduction

In tribology, when friction pairs are brought together to slide relative with one another under a non-zero normal force, the initial wear process is referred to as running-in [1]. In the running-in process, the mating surfaces of the friction pair gradually become adequate fitting. Studies [2-3] have proved that the running-in has an important influence on improving the performance and service life of the friction pair. Therefore, monitoring and identifying the running-in process of the friction pair are significant to improve the reliability and the economy of the equipment [4].

The running-in process of friction pair can be monitored through the tribological characteristics, such as surface morphology [5], wear rate [6], oil analysis [7], friction signal [8], etc. However, it is challenging to directly obtain the surface morphology and wear rate of friction pairs in real-time during the equipment operation. And oil analysis is mostly used for monitoring the equipment off-line regularly at present. Friction signals, including the friction-induced vibration (*FIV*) [9], the friction coefficient [10], the friction heat [11], the friction force [12], can be collected online without interrupting the operation of mechanical equipment and characterizing the wear state. Therefore, the friction signal is an ideal means to monitor the running-in process. Due to the *FIV* signal containing lots of information on a friction-wear process and can be collected easily, the *FIV* signal is suitable for monitoring the running-in process.

According to the reference [13-14], two types of *FIV* signals can be generated in a friction-wear process. One is the *FIV* signal with a large amplitude and lumping on a narrow frequency band (hereinafter referred to as the periodic *FIV*), such as the brake squeal [15]. The other is the *FIV* signal with small amplitudes and distributes in a wide frequency band (henceforth called the aperiodic *FIV*), usually referred to as surface or roughness noise [16]. Scholars have conducted extensive studies on monitoring the running-in process by the aperiodic *FIV* signal. For example, Ding et al. [17] found that the friction noise attractor is chaotic, and the attractor evolvement of friction noise can help identify the wear process. Zhou et al. [18] studied the correlation between the aperiodic *FIV* and friction coefficient under different friction states by cross recurrence plot. Xu et al.[19] characterize the surface wear state of a sliding-rolling contact through the features of the aperiodic *FIV*. Yu et al. [20] identify the friction state through the frequency band energy of the aperiodic *FIV*. Thus, the wear process can be characterized by the aperiodic *FIV*. However, due to the aperiodic *FIV* distributes in a band frequency range, its amplitude variation must be analyzed by the wear experiments to investigate whether the extracted vibration signal could indicate the friction behavior, which is problematic in a practical application.

Fortunately, due to the frequency of periodic *FIV* being close to the system's natural frequency, it can be identified effectively [21]. Scholars have proved that normal load, surface roughness, sliding velocity, etc., have an important influence on the *FIV* with periodic characteristics [22-25], which indicates that the periodic *FIV* is suitable for wear state evaluation. However, the periodic *FIV* is so weak that it often submerges in background noise under oil lubrication conditions. Therefore, the periodic *FIV* under oil lubrication has not received the attention that would have been deserved [26-27]. As a result, the periodic *FIV* characteristics in the running-in process are still unrevealed, and further investigation is still needed.

In this paper, a running-in experiment of ball-disk friction pair was carried out on a test rig to fulfill this goal. The particular focus is put on extracting the periodic *FIV* under oil lubrication and studying its evolvement process and transition mechanism in the running-in process. This work is organized as follows. Firstly, the experimental details are illustrated in Section 2. Then, the workflow of the experiment and the data processing is elaborated briefly in Section 3. Then in Section 4, the extraction methods of the periodic *FIV* signals using the harmonic wavelet packet transform (*HWPT*) are elaborated detailedly; and the evolvement process and transition mechanism of the periodic *FIV* in different wear stages are investigated respectively by the root mean square (*RMS*) and the composite surface roughness. In the end, the conclusions are listed in Section 5.

## 2. Experimental details

### 2.1 Experimental equipment

The ball-on-disk running-in experiments were conducted on a friction-abrasion testing machine (model CFT-I, Zhongkekaihua Company), as illustrated in Fig. 1. In Fig.1 (a), an electrical motor with rotational speed from 0-2000r/min is used to drive the movable bench reciprocating in the stroke of 1-25mm through an eccentric mechanism. The disk specimen is mounted on a movable bench firmly by a hold-down plate and moves together. The ball specimen is installed on the load device and remains stationary in the experiment. In the experiment, the ball specimen is pressed on the disk specimen by the load device. The load pressed on the ball specimen can be regulated from 10 to 200N. The friction force between the ball and disk specimen is measured by the force sensor, processed by the data acquisition system of the testing machine, and finally recorded in the form of friction coefficient. A tri-axial accelerometer (model 356 A16 ICP, PCB Piezotronics Company) with a sensitivity of 10mV/g and a range of ±50g is fixed on the disk specimen lower surface to measure the vibration signal. A signal collecting device (INV-3062T2, China Orient Institute of Noise & Vibration) stores the vibration signals into the computer. The surface roughness of the ball-disk friction pair is measured by a confocal laser scanning microscope (OLS4000, Olympus Corporation).

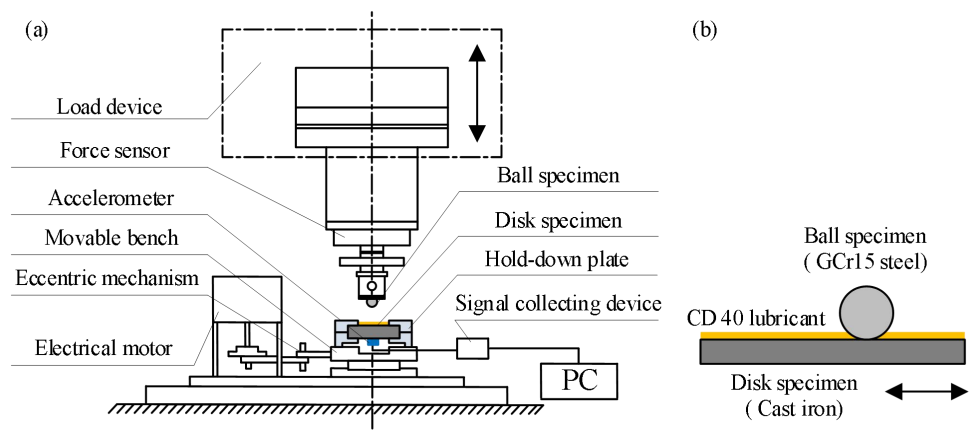

**Fig. 1.** Schematic diagram of the testing machine：(a) testing machine (b) friction pair

**2.2 Friction pairs**

The ball and disk specimens are used as the friction pair for the running-in experiment, as depicted in Fig. 1(b). The ball specimen is constructed of GCr15 steel with Φ6mm diameter and hardness of 750HV. The disk specimen is made up of cast iron with Φ16mm diameter and hardness of 480HV.

## 2.3 Experiment method

### 2.3.1 Running-in experiment under lubrication

The drop lubricated running-in experiments were carried out under ambient condition (293 K, 45% relative humidity). The lubricant was CD 40 lubricating oil, an ordinary marine lubricant. The amount of lubricating oil used in the experiment was 200ml. In the testing, the ball specimen was pressed on the upper surface of the disk specimen under a load of 100N and stayed stationary. The disk specimen was driven to reciprocate in the stoke of 5mm by the motor with a rotational speed of 400 r/min. The calculated Hertz contact stress between the friction pair was 3Gpa, and a calculated relative sliding velocity between the two was 0.067m/s. Three running-in wear tests were carried out to ensure the repeatability. Each test lasted for 60 minutes and the averages were calculated.

(1) Setting the wear stage

According to the friction coefficient variation trend, the wear stage of the friction pair was set [28]. Specifically, when the friction pair is in a running-in wear stage, the friction coefficient decreases gradually; when the friction coefficient presents a smooth and stable variation, the friction pair is in a stable wear stage. Fig. 2 depicts the friction coefficient variation in the running-in experiment. The fitting curve displays that the friction coefficient has a more significant value of 0.129 at the initial moment and then decreases to 0.103 gradually in the first 40 minutes. Finally, the friction coefficient fluctuates around 0.103 from 40th to 60th min. Therefore, to facilitate the analysis, the first 40 minutes of the experiment were set as a running-in wear stage, and the 40-60 minutes were set as a stable wear stage.

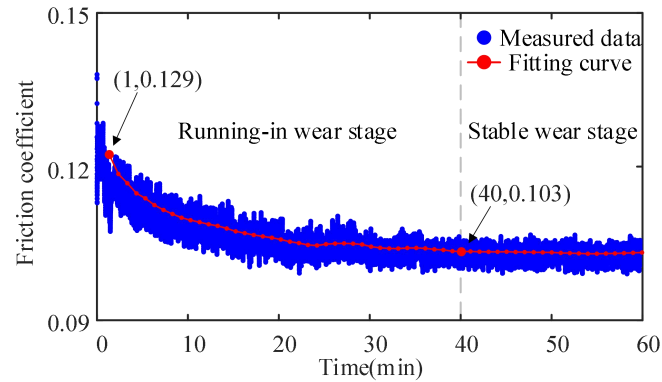

**Fig. 2.** Friction coefficient variation in the running-in experiment

(2) Collecting vibration signals

A signal collecting device was applied to collect the vibration signals every 6 seconds with a sampling interval of 0.039ms and 10240 sampling points. In the running-in experiment, 10 groups of acceleration signals were collected every minute.

(3) Measuring surface roughness

The surface roughness of the ball and disk specimen was measured by the confocal laser scanning microscope before and after the running-in experiment. A total of 4 groups of surface roughness was obtained. As shown in Fig. 3, before the running-in experiment, the surface roughness of the ball specimen ($Sa_{ball-1}$) is 0.124μm, and that of the disk specimen ($Sa_{disk-1}$) is 0.547μm; after the experiment, the surface roughness of the ball specimen ($Sa_{ball-2}$) is 0.554μm, and that of the disk specimen ($Sa_{disk-2}$) is 0.279μm.

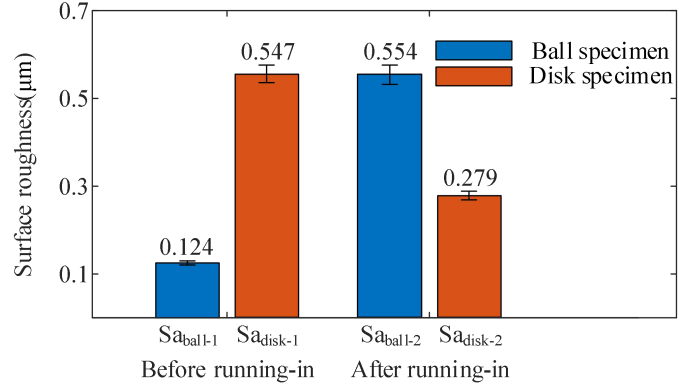

**Fig. 3.** The surface roughness of friction pair

**2.3.2 Experiments with squeal under oil-free condition**

To determine the frequency distribution of periodic *FIV*, a 5-minutes experiment that can generate squeal noise was conducted under an oil-free condition. In the testing, a test load of 20N was applied to the ball specimen, and other conditions are consistent with the lubricated running-in experiment. The initial surface roughness of the ball-disk friction pair is the same as that under the lubricated running-in experiment. Only the squeal signals were collected in the experiment.

**2.4 Determination of lubrication state**

The lubrication state was determined by the film thickness ratio $\lambda$ [29], that is, the fluid lubrication ($\lambda>3$), the mixed lubrication ($3>\lambda>1$), and the boundary lubrication ($\lambda<1$). $\lambda$ is defined as follows:

$$\lambda = \frac{h_{min}}{\sigma_c} \tag{1}$$

In Eq. (1), $\sigma_c$ is the composite roughness of the two contact surfaces, which can be calculated from Eq. (2). $h_{min}$ is the minimum lubricant film thickness according to Hamrock-Dowson formulation [30-31] given in Eq. (2) - (4).

$$\sigma_c = \sqrt{\sigma_1^2 + \sigma_2^2} \tag{2}$$

$$h_{\min} = 7.43R\left(1-0.85e^{-0.31k}\right)\left(\eta\mu/E^*R\right)^{0.65}\left(P/E^*R^2\right)^{-0.21} \tag{3}$$

where $\sigma_1$ is the surface roughness of the ball specimen; $\sigma_2$ is the surface roughness of the disk specimen; $R$ is the composite radius; $E^*$ is the composite elasticity modulus. The two parameters can be calculated from Eq. (3) and Eq. (4), respectively. The other parameter values of Eq. (3) -(5) are shown in Table 1.

$$\frac{1}{R} = \frac{1}{R_a} + \frac{1}{R_b} \tag{4}$$

$$E^* = \frac{2}{\frac{1-v_a^2}{E_a} + \frac{1-v_b^2}{E_b}} \tag{5}$$

Table 1 The parameter values

| Parameter | Physical meaning | Values |
|---|---|---|
| $k$ | the ellipticity parameter [32] | 1 |
| $\eta$ | the absolute viscosity of the lubricant | 0.139 Pa·s |
| $\mu$ | the mean velocity between the friction pair | 0.0333 m/s |
| $P$ | the normal load | 100 N |
| $R_a$ | the radius of ball specimen | 3 mm |
| $R_b$ | the radius of the disk specimen | Infinite |
| $v_a$ | the Poisson's ratio of steel ball | 0.3 |
| $v_b$ | the Poisson's ratio of the disk | 0.269 |
| $E_a$ | the elasticity modulus of the ball | 208 GPa |
| $E_b$ | the elasticity modulus of the disk | 209 GPa |

Based on the above data, the lubricant film thickness is 5.51nm under a load of 100N. The film thickness ratio is 0.0098 before the experiment and 0.0089 after the experiment, respectively, that is, λ<1. Therefore, the running-in experiment was under the boundary lubrication state.

## 3. Workflow

The workflow for the experimental study is depicted in Fig. 4. Two types of experiments are conducted. For the experiment under oil-free conditions, only the squeal signal is sampled. For the lubricated running-in experiment, the surface roughness, friction coefficient, and vibration signals are recorded. The power spectrum analysis is then conducted to study the periodic component in the measured signals. By comparing the frequency components of the squeal signal and the vibration signal, the periodic *FIV* submerged in the vibration signal of the lubricated running-in experiment is identified. Then, the periodic *FIV* is extracted using the harmonic wavelet packet transform (*HWPT*) method [33]. Following the extraction, the evolvement process of the periodic *FIV* has performed analysis by the *RMS*. Finally, the evolution mechanism of periodic *FIV* is discussed with the help of composite surface roughness.

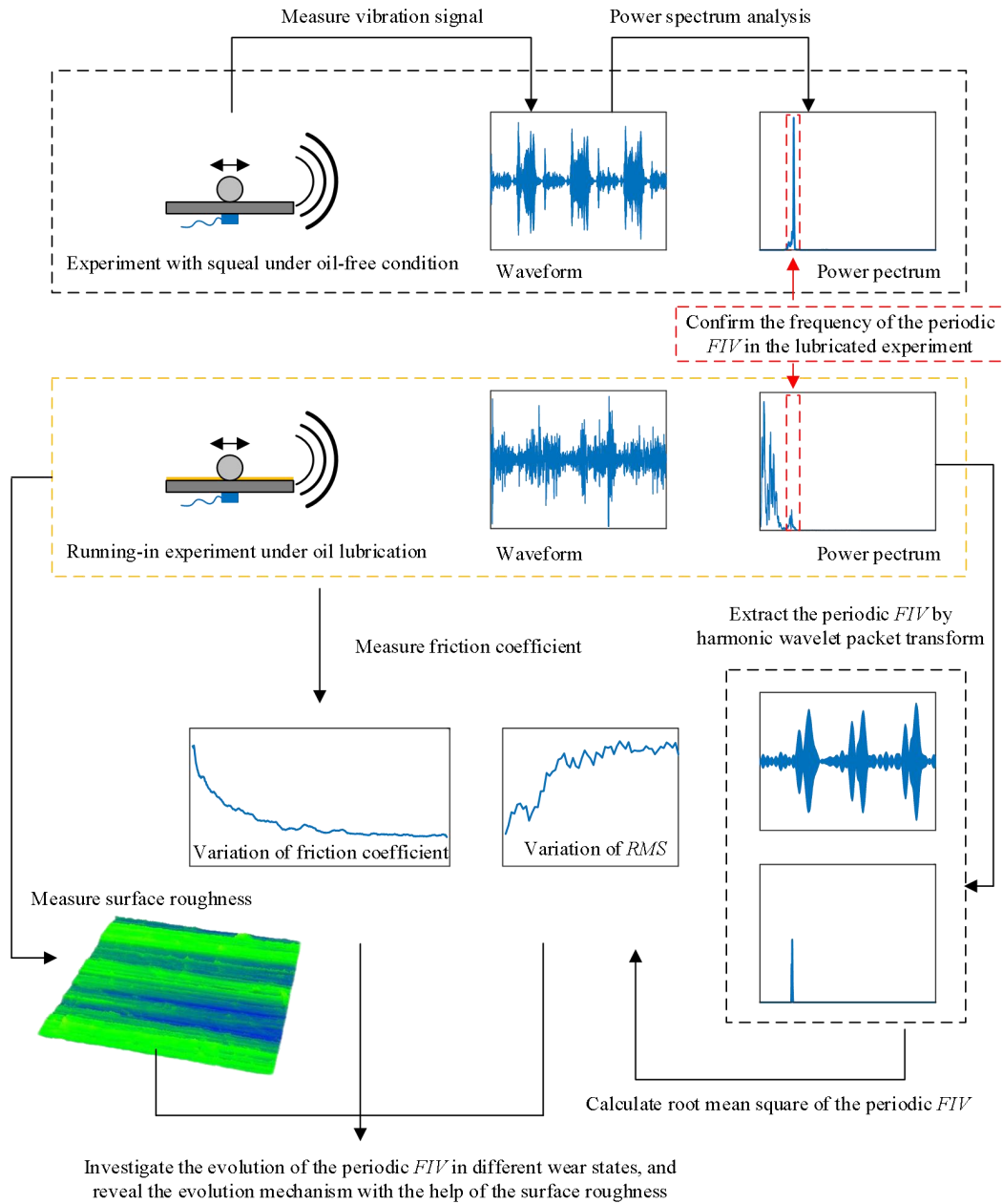

**Fig. 4.** The workflow for experimental study

## 4. Results and discussion

### 4.1 Extracting the periodic FIV signal

#### 4.1.1 Time-frequency analysis of the original vibration signals

The power spectrum analysis was performed to investigate the periodic components in the original signal better. Fig. 5 depicts the waveform and the power spectrum of the original vibration signals at different moments. A log scale is used for the frequency axis to show frequency distribution more clearly in Fig. 5(b). Fig. 5(a) displays the waveform of the original signals presents aperiodic variation. However, Fig. 5(b) shows multiple prominent periodic components in the original signal, such as 319Hz, 853Hz, and 2332Hz. Therefore, there are various periodic components in the measured vibration signals; and the periodic *FIV* is submerged in the original signal.

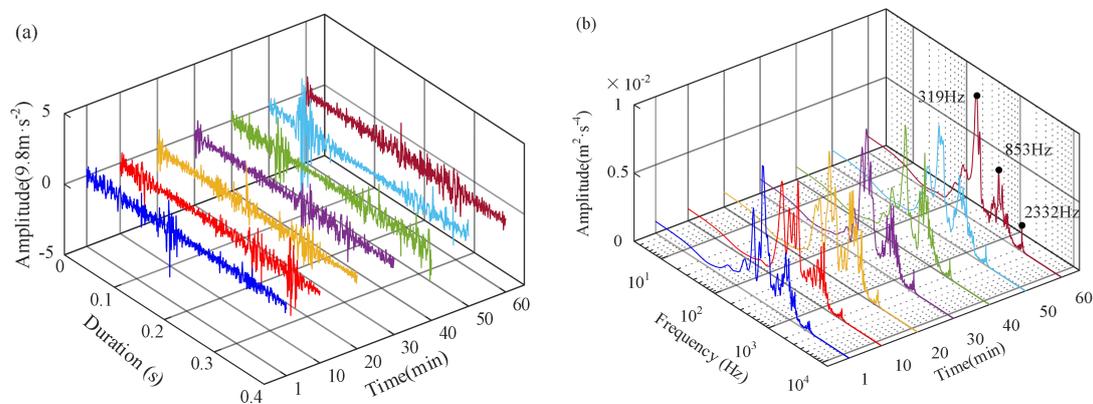

**Fig. 5.** The waveform and power spectrum of original signals under lubricated condition at different moments: (a) the time-domain waveform (b) the power spectrum

### 4.1.2 Identifying the frequency of periodic *FIV*

The power spectrum analysis was performed on the squeal signal collected under oil-free conditions to identify the frequency of the *FIV*. Fig. 6(a) depicts that the waveform of the squeal signals presents periodic variation obviously; Fig. 6(b) shows that the squeal frequency is 2325Hz, 2412Hz, 2381Hz, 2384Hz, and 2425Hz, respectively, with an average of 2385Hz. According to the hammering model proposed by Rhee [34], the *FIV* is caused by the hammering action of the asperities between the contact surface similar to that in the modal test, leading to the dominant frequency of the *FIV* is often consistent with the natural frequency of the tribo-system

[21,35-36]. Therefore, the natural frequency of the tribo-system is close to 2385Hz. Correspondingly, the frequency around 2385Hz may be the primary frequency component of the periodic *FIV* in the lubricated running-in experiment, as shown in Fig. 5(b).

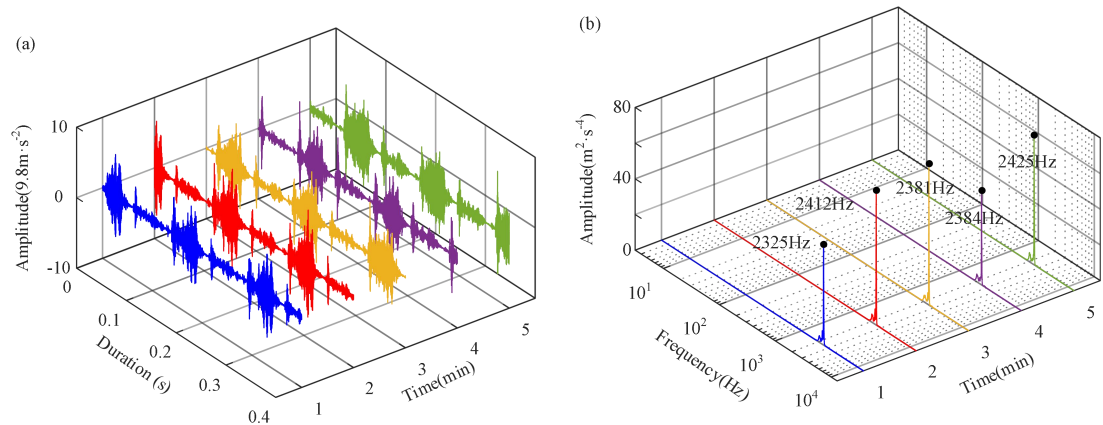

**Fig. 6.** The waveform and power spectrum of squeal signals under dry friction at different moments: (a) the time-domain waveform (b) the power spectrum

**4.1.3 Features of the extracted periodic FIV**

The *HWPT* was carried to extract the high-frequency periodic component from the original signal measured in the lubricated running-in experiment to reduce the noise. Fig. 7 depicts the distribution of the high-frequency periodic component and its amplitude. As illustrated in Fig. 7, the periodic component mainly distributes from 2300Hz to 2400Hz, and there is not much difference in the frequency among different groups of signals. Thus, the components in the frequency band from 2300Hz to 2400Hz were extracted using *HWPT* to reconstruct as the periodic *FIV* signal. More specifically, a 7-level HWPT was carried to decompose the original signals into 128 bands with a width of 100Hz, and the 23rd frequency band was extracted and reconstructed directly.

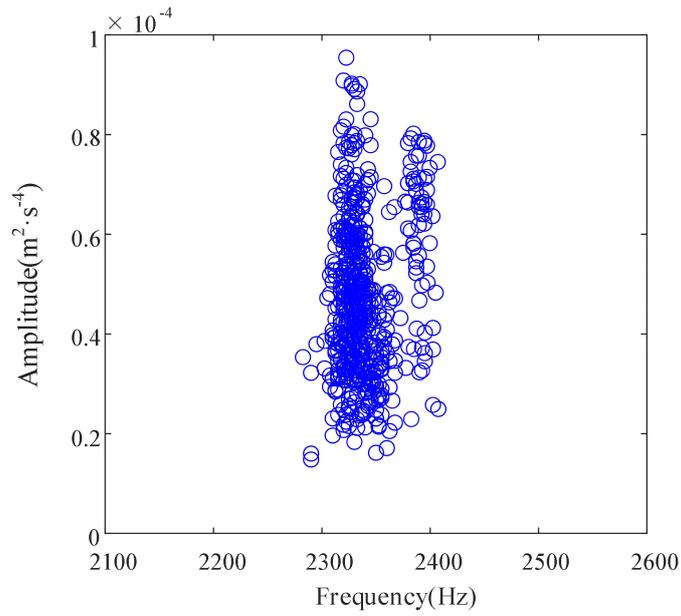

**Fig. 7.** The distribution of high-frequency periodic component

The waveform and power spectrum of the extracted periodic *FIV* signals are shown in Fig. 8. It can be seen from Fig. 8(a)-(b) that the waveform of the extracted *FIV* signals displays periodic characteristics. Furthermore, the amplitudes of the extracted periodic *FIV* signals are distinct at different moments in the wear experiment, i.e., the amplitude increases gradually at first and then trends to be stable, indicating that the extracted periodic *FIV* signal can distinguish the wear state.

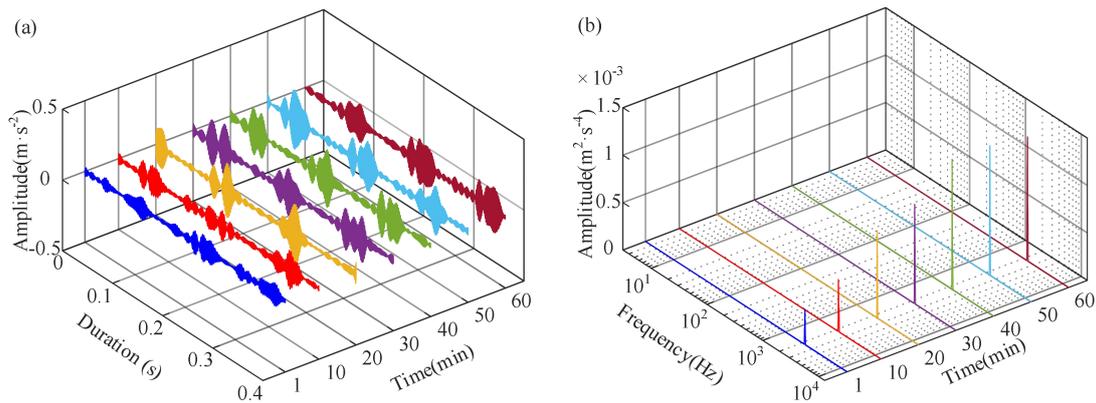

**Fig. 8.** The waveform and power spectrum of the FIV signals under lubricated condition at different moments: (a) the time-domain waveform (b) the power spectrum

### 4.2 Evolvement of the periodic FIV

#### 4.2.1 RMS variation law

The *RMS* of the periodic *FIV* was calculated, and it displays preferable regularity in different wear states, as shown in Fig. 9. Within the first 40min of the experiment, the *RMS* value is small and reveals an upward trend; After the 40th minute, the *RMS* value is large and fluctuates steadily. According to the experiment condition, the friction pair has also experienced the running-in wear stage (0-40 minutes) and steady wear stage (40-60 minutes), indicating that the extracted periodic *FIV* is closely related to the wear stage.

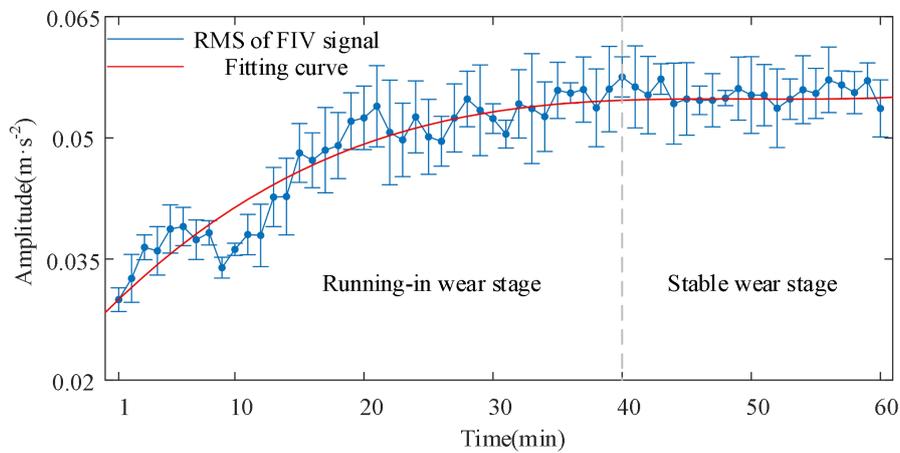

**Fig. 9.** *RMS* value variation of the extracted periodic *FIV* signal in the running-in experiment

**4.2.2 Evolvement mechanism**

As described in section 2.4, the running-in experiment was conducted under boundary friction, indicating that the lubricant film was not thick enough to separate the friction surface [37]. Thus, the load was entirely undertaken by the asperities between the friction pair rather than by the lubricant film, demonstrating that the friction pair was in a strong contact condition [38]. Correspondingly, the periodic *FIV* was triggered by the hammering action of the asperities between the contact surface [34]. Moreover, the film thickness ratio remained stable in the experiment, as calculated from equation (3)-(5), which indicates that the contact condition between the friction pair was mainly affected by the surface roughness. As depicted in Fig. 3, the surface roughness of the disk and ball samples shows an opposite variation trend in the test process. Accordingly, it is not comprehensive to study the evolvement mechanism of the periodic *FIV* in the wear process only from the surface roughness of a single sample. Therefore, the evolvement mechanism of the periodic *FIV* is investigated by the composite surface roughness of the friction pair. The specific analysis is as follows.

(1) Running-in stage (0-40min)

The initial composite roughness of the friction pair is 0.56μm at the beginning of the experiment as calculated from equation (2). According to Fig.3, the surface roughness of the ball specimen is increased by 0.43μm after the experiment, while that of the disk specimen is reduced by 0.268μm, indicating the roughness increase of the ball specimen is greater than the roughness decrease of the disk specimen in the experiment. Thus, the composite roughness increases in the running-in stage. Studies have shown that high surface roughness can reduce the effective contact area [39-40]. Therefore, the contact is strengthened progressively under a constant load as the decrease in contact area. With the strength of the contact, the hammering action of the asperities is also enhanced. As a result, the intensity of the periodic *FIV* triggered by the contact is also progressively enlarged. Correspondingly, the *RMS* value of the extracted periodic *FIV* signals is small initially and displays an ascending with the running-in going on, as shown in Fig. 9.

(2) Stable wear stage (40-60min)

The friction pair entered the stable wear stage from the 40th-60th minute. In the stable wear stage, the surface roughness of the friction pair remains steady until the end of the experiment. Accordingly, the surface roughness of the ball and disk specimens fluctuated around 0.554μm and 0.279μm separately at the stable wear stage, as depicted in Fig. 3. Thus, the composite roughness was stable at around 0.62μm in this stage such that the effective contact area also persisted stable. Therefore, the contact between the friction surface was stable. Under a stable contact, the hammering action of the asperities is also stable. Correspondingly, the *RMS* value of the extracted periodic *FIV* signals displays a stable trend, as shown in Fig. 9.

## 5. Conclusion

In the present work, a running-in experiment of ball-disk friction pair was conducted under lubrication. The *HWPT* method is used to extract the periodic *FIV* effectively from the measured vibration signal. The evolvement process of the periodic *FIV* was investigated by *RMS* value in the running-in and stable wear stage. Furthermore, the evolution mechanism was discussed with the help of the composite surface roughness. The following conclusions can be achieved.

(1) The periodic *FIV* can be generated under lubrication, whose frequency of the periodic *FIV* is nearly stable. Moreover, the periodic *FIV* is weak with low amplitude and often submerged in the background noise. The *HWPT* is an effective method to extract the periodic *FIV* in a running-in process.

(2) The *RMS* increases progressively in the running-in stage, and then remains stable fluctuation in the stable wear stage, which follows the same variation trend as the composite surface roughness of the friction pair. Therefore, the *RMS* of the periodic *FIV* signal can be applied to identify the wear stage of a friction pair in practice.

In this research, a severe wear stage is not included. In the future, we will conduct further study on the periodic *FIV* in a whole wear process under oil lubrication.

**Acknowledgments**

We appreciate each of the reviewers and the anonymous reviewers for their valuable comments and suggestions for improving the quality of this paper. This present project is supported by the National Natural Science Foundation of China (Grant No. 51879020 and 51679022), the "double first-class" construction project (innovative project) of Dalian Maritime University (Grant No. BSCXXM006), and the Natural Science Foundation of Fujian Province, China (Grant No. 2018J01498).